# Harnessing Abundance: A Generativity Perspective on Human-GenAI Collaboration

*Full Paper*

**Yoram M Kalman**
The Open University of Israel
yoramka@openu.ac.il

**Yun Wan**
University of Houston Downtown
wany@uhd.edu

## Abstract

Research on human-GenAI collaboration yields conflicting findings: GenAI can enhance creativity yet reduce collective diversity, with uneven benefits across skill levels. Rather than treating these as contradictions, we argue they reflect a core feature of GenAI: abundance. GenAI makes ideas, drafts, and recombinations plentiful, potentially expanding the hypothesis space and surfacing unanticipated possibilities. However, abundance alone doesn't ensure better outcomes. We propose generative fit as a unifying mechanism explaining when abundance yields productive creativity and when it backfires. Drawing on Generativity Theory, generative fit captures how well a system's generative potential complements a community's generative capacities. We develop a conceptual framework for collaborative human–GenAI settings where participants share goals, depend on one another, and must integrate diverse contributions. By mapping abundance to cognitive, social, and organizational factors of collective creativity, we explain apparent tradeoffs and offer actionable implications for designing workflows that convert abundance into valued creative outcomes.

## Introduction

Generative AI (GenAI) has rapidly moved from a novel tool to a routine collaborator in creative and knowledge work. Individuals and teams now use GenAI to brainstorm, draft, critique, design, and explore alternatives at a speed and scale that were previously impractical. In this conceptual paper we focus on settings where collaborators share goals, depend on each other's inputs, and must integrate diverse contributions into a coherent joint output. Most studies on human-GenAI creative collaboration occur in such settings, and what matters in such collaborations is not only whether GenAI generates "good ideas," but how its contributions shape collective exploration, coordination, and integration.

In this context, the empirical evidence is best described as *mixed* rather than uniformly positive or negative. Many studies show trade-offs within the same setting: GenAI can improve one outcome while worsening another, or help some users and hinder others depending on collaboration modality and expertise. Across tasks such as creative writing, marketing, and innovation activities, GenAI often improves individual performance or perceived creativity, yet these gains can coincide with collective downsides. Doshi and Hauser (2024), for example, report higher individual creativity alongside reduced collective diversity of outputs. Mixed patterns also appear once study results are broken down by expertise and use mode: Chen and Chan (2024) show that feedback-oriented use benefits non-experts, while ghostwriting can hinder experts and is associated with higher similarity among outputs; in contrast, Jia et al. (2024) report skill-biased creativity gains that accrue more to higher-skill employees. Taken together, the evidence points to contingent trade-offs, not a single "GenAI effect", and it motivates a framework that explains *when* and *why* GenAI produces different patterns in collaborative creative work.

We argue the field is missing a unifying lens because it has not fully theorized GenAI's defining contribution to creative collaboration: abundance. GenAI expands the hypothesis space and enables emergent affordances and serendipitous combinations.

We develop a conceptual framework grounded in generativity theory with emphasis on the concept of generative fit. Generative fit was defined as the extent to which the generative potential of an information system complements and enhances the generative capacities of a community (Avital & Te'Eni, 2009), and later work emphasizes generative outcomes as the product of combinatorial innovation emerging from the match between architecture and community (Acar et al., 2024) . Building on this foundation, our framework contributes to generativity theory in three ways: it identifies *abundance* as GenAI's distinctive generative affordance and differentiates it from the resource abundance of prior technologies such as search engines; it reframes *generative fit* as a design-time configuration variable that shapes collaboration outcomes rather than a post-hoc reconstruction of them; and it explains an *abundance-to-monoculture* paradox as a predictable consequence of insufficient fit under bottleneck-shifted conditions.

# Theoretical Foundations

## *Collective creativity and innovation: why collaboration factors matter*

Acar et al.'s (2024) interdisciplinary review of collective creativity integrates research across domains and organizes key antecedents of collective creativity into cognitive, social, and organizational "architectures". Acar et al. (2024) also distinguish between attention-based, divergence-based and convergence-based teams (collectivity types). In our analysis we focus on convergence-based teams – settings where collaborators share goals and their search and development efforts are interdependent. Most studies on human-GenAI creative collaboration occur in such settings, and they impose unique coordination and integration demands. In such settings, creativity does not primarily come from "having more ideas," but from how diversity is integrated, how conflict is handled, how trust is built, and how coordination enables iterative convergence on a shared artifact. We will note here that even individual "co-creation" episodes in which individuals use GenAI are a form of convergence-based human-GenAI collaboration.

This matters for GenAI because most GenAI deployments are not simply "a smarter tool"; they introduce a new actor capable of injecting abundant contributions into the collective process, often faster than the human members of the team can absorb, evaluate, and integrate. Abundance changes the relative importance of collaboration factors. For instance, if idea generation becomes trivial, then evaluation, integration, and coordination become bottlenecks. Conversely, if coordination is partially shifted into the tool (e.g., asynchronous AI-augmented ideation), coordination overhead can decline in certain tasks, but only if the workflow design fits the task.

## *Generativity and generative fit*

Generativity theory is an Information Systems theoretical framework that explains how systems (e.g., platforms and infrastructures) produce novel, unanticipated outcomes by enabling recombination and participation. A core implication of this theory is that outcomes are not determined solely by technology capability; they emerge from the relationship between the generative architecture and the generative community–their match is termed generative fit (Thomas & Tee, 2021). Avital and Te'eni (2009) conceptualize generative fit as the extent to which the system's generative potential complements and enhances a community's generative capacities, and they explicitly frame generative fit as a design-oriented construct: designers can build features that help users leverage generative potential and overcome negative unintended consequences. Furthermore, Avital & Te'eni (2009) decompose generative fit into evocative fit (supporting ideation and exploration), adaptive fit (supporting flexible task-technology alignment), and open-ended fit (supporting emergence and unanticipated outcomes).

## *Extending generative fit to human-GenAI collaboration*

We propose applying the concept of generative fit to analyzing the role of GenAI in human-GenAI collaboration. First, it allows us to theorize how abundance gets "converted" into outcomes (not merely whether GenAI is present). Second, it anticipates that fit is not "one size fits all": the fit needed for a novice brainstorming task may be different from the fit needed for expert decision support, the fit needed to enhance individual creativity might be different from the fit needed to create collective diversity, and so on.

We now turn to the central theoretical contribution of this paper: the abundance principle. In the preceding section, we established generative fit as the mechanism by which GenAI's capabilities translate, or fail to translate, into productive outcomes. But this raises the central question: what exactly is GenAI contributing to the collaboration that requires such fit? The answer, we argue, is abundance. Where previous information systems augmented specific human capacities (calculation, retrieval, communication), GenAI is distinctive in its capacity to generate plentiful contributions across virtually every dimension of creative work: ideas, critiques, drafts, perspectives, recombinations. The generation is often faster than collaborators can absorb them. This abundance is not merely quantitative; it completely reshapes the collaboration landscape by shifting bottlenecks from generation to evaluation, from ideation to integration. Understanding abundance as GenAI's defining contribution allows us to explain why the same technology produces divergent outcomes across contexts: abundance interacts differently with cognitive, social, and organizational factors depending on the degree of generative fit. In the next section, we develop the abundance principle in detail and show how it generates the apparent paradoxes observed in empirical research.

## The abundance principle: what it is, and why it produces paradoxes

### *The abundance afforded by GenAI*

We define abundance as GenAI's general-purpose capacity (E. K. Chen et al., 2026) to enhance collective creativity and innovation by simultaneously supporting variation (the low-cost generation of diverse inputs) and integration (the synthesis of those inputs into novel outcomes), with breadth across creative processes, depth within each, and freedom from human cognitive constraints. These contributions can include ideas, data and information, alternative framings, partial artifacts, critiques, and even process-level actions such as planning, decomposition, delegation, and in agentic systems, autonomous subtask execution. Two points of differentiation warrant emphasis. First, unlike search engines or databases, which deliver abundant *retrieval* of pre-existing content in response to queries, GenAI produces abundant *recombination and novel synthesis*, generating artifacts that did not previously exist. Second, abundance at the system level does not translate uniformly into abundance at the individual level: a user's realized abundance is bounded by domain knowledge, prompting skill, and the social and organizational structures in which the interaction is embedded, raising distributional concerns that our framework treats as a core object of study.

Abundance is therefore multidimensional. It can take the form of idea abundance (many candidate concepts and alternatives), perspective abundance (multiple framings, personas, and role-based interpretations), draft abundance (rapid production of multiple artifacts such as text, code, or images), feedback abundance (continuous critique, evaluation criteria, and revision suggestions, often functioning as a "second mind" in prewriting and refinement), and process abundance (support not only for content generation but for workflow orchestration and stepwise execution).

Abundance is tied to emergence. GenAI's scale and training produce emergent capabilities and unprogrammed affordances; users routinely uncover new uses and behaviors (Holtzman et al., 2025). In generativity terms, abundance enlarges the space of recombinations and trajectories. Abundance reduces participation barriers. Natural language interfaces lower the threshold for non-programmers and domain experts to participate in innovation activities. Thus, GenAI also increases the number and diversity of human participants who can contribute to the collaborations.

### *How can abundance lead to monoculture?*

If abundance is real, why do we observe convergence toward the same high-probability solutions, with reduced diversity as the result? This is the apparent paradox: more abundance can lead to less diversity, not because variety is unavailable, but because selection concentrates it. Because abundance can produce monoculture. We borrow the term monoculture from agriculture. Monoculture often emerges when many cultivars exist, but agricultural scientists and breeders develop one standardized variety by selecting and recombining desirable traits from across that diversity, and then deploy it at scale, reducing biodiversity (Altieri, 1999). In human-GenAI collaboration, a parallel narrowing can emerge from human cognitive and attention constraints combined with AI defaults that concentrate choice around high-probability outputs. The result is a loss of idea diversity: a narrow pattern space of individually plausible yet closely clustered ideas that parallels biodiversity loss in agricultural monoculture.

The mechanism for leading to monoculture is that abundance shifts the bottleneck. When GenAI can generate far more inputs than humans (or teams) can evaluate and integrate, the limiting resources become attention, selection, and integration capacity. As is well known from research on information overload, when the volume of potentially relevant inputs exceeds the time and cognitive capacity available to process them, additional information becomes a hindrance rather than a benefit (Schick et al., 1990). In other words, GenAI abundance shifts the constraint from generating options to filtering, evaluating, and integrating them, exactly the conditions under which overload research predicts shallow processing and degraded judgment unless filtering mechanisms are in place (Eppler & Mengis, 2004). Under these conditions, teams may fall back on heuristics that compress the search space. Overload conditions trigger coping responses like simplifying heuristics, satisficing, and reliance on salient or default options, which reduces exploration of the full option space. At scale, those coping responses become a selection regime: when evaluation capacity is scarce, people repeatedly pick the most legible and easily justified outputs, even when many alternatives exist (Bawden & Robinson, 2020). That is one pathway from overload to monoculture, and when using GenAI this can show up as predictable coping moves: accepting the first plausible output, reusing the same prompt template, relying on “ghostwriting” rather than interactive co-creation, or converging prematurely to reduce coordination overhead. To fully leverage the abundance of GenAI and, meanwhile, mitigate the side effect of monoculture, we suggest the alignment between the configurations of human-GenAI collaboration specs and the tasks, or the fitness calibration through generative fit framework (Avital & Te’Eni, 2009).

## Generative fit: the calibration of human-GenAI creative collaboration

A generative fit view of human-GenAI collaboration frames collective creativity as a two-part challenge: assembling a sufficiently diverse pool of knowledge and perspectives, and then synthesizing or integrating that diversity into a coherent shared solution. Building on Acar et al.'s (2024) architecture-based account of the antecedents of collective creativity and innovation, we argue that GenAI-driven abundance can strengthen or weaken these antecedents depending on the form of generative fit it enables: evocative fit – supporting ideation and exploration, adaptive fit – supporting flexible task-technology alignment, or open-ended fit – supporting emergence and unanticipated outcomes (Avital & Te’Eni, 2009). The same abundance that expands the space of options can also shift the bottleneck to attention, selection, and integration, so that outcomes hinge on whether fit supports exploration, flexible alignment to task and expertise, and the emergence of unanticipated directions. We therefore use the cognitive, social, and organizational architectures as a simple organizing structure for showing how GenAI can either amplify or impede collaboration, depending on how abundance is configured and governed.

### *Cognitive architecture*

At the cognitive level, collaborative work benefits from demographic, geographic, and informational diversity because these expand the space of candidate solutions.

Abundance can support evocative fit when GenAI outputs are used to stimulate exploration rather than to supply a single “best” answer. In practice, this means producing multiple candidate hypotheses, analogies, sketches, or reframings that trigger new human associations and encourage parallel lines of inquiry. Abundance can support adaptive fit when teams match support to task demands and expertise, for example by shifting from broad exploration to focused elaboration, or by offering different levels of scaffolding to novices versus experts. Abundance can support open-ended fit when it preserves the possibility of surprise and ongoing change, allowing new directions to emerge through recombination over time, instead of locking the team into an early framing. Without these forms of fit, abundance can overwhelm sensemaking: teams face too many options to evaluate, fall back on shallow selection strategies, and converge prematurely, reducing realized cognitive diversity even when the upstream pool is large.

### *Social architecture*

At the social level, Acar et al. (2024) highlight antecedents that govern whether diverse inputs can actually be integrated, including collaborative dynamics, competition, trust, evaluation, feedback and criticism, network position and structure, and outside tie characteristics.

Abundance can support evocative fit socially when it provides shared material for joint interpretation and constructive disagreement, strengthening collaborative dynamics and enabling critique to operate as a creativity resource rather than a coordination burden or source of mistrust. It can support adaptive fit when teams can regulate the intensity and timing of GenAI input, for example by decoupling generation from evaluation, or by varying who uses GenAI and when, so that trust and social influence are calibrated rather than flattened. It can support open-ended fit when network structures and outside ties continue to introduce genuinely distinct information, preventing the collective from repeatedly rediscovering the same high-probability patterns. When these fits are absent, abundance can intensify competition for attention, introduce opaque "answers" that weaken trust, and accelerate social convergence through repeated exposure to similar suggestions.

### *Organizational architecture*

At the organizational level, Acar et al. (2024) point to formal mechanisms that shape collaborative creativity: HRM practices, incentives, state policies, organizational structure, collaboration and innovation tools, and leaders' characteristics and behavior.

These mechanisms determine whether the collective has the capacity and permission to do the work that abundance demands. Abundance can support evocative fit when organizations legitimize exploration, allocate time for divergent search, and equip teams with tools that make alternatives comparable rather than merely plentiful. It can support adaptive fit when governance and workflow design allow teams to adjust GenAI use by task phase, uncertainty, and expertise, rather than enforcing one uniform pattern. It can support open-ended fit when structures and leadership protect long-run generative capacity, for example by preserving boundary spanning and learning loops, rather than replacing them with standardized GenAI outputs. Without these forms of fit, organizational adoption of GenAI could tend to convert abundance into efficiency-oriented standardization, narrowing the range of acceptable options, weakening exploratory learning loops, and ultimately producing convergence that is fast but fragile.

In all these ways, as well as others, generative fit is the mechanism that determines whether abundance becomes a creativity amplifier or a monoculture engine in innovation.

## Illustrative Scenarios

To demonstrate how the generative fit framework applies in practice, we analyze four recent empirical studies of human-AI collaboration through the lens of abundance, generative fit, and the cognitive, social, and organizational architectures outlined above. Each scenario identifies instances of good and poor fit within the same study, showing how identical AI deployments produce divergent outcomes depending on whether evocative, adaptive, and open-ended fit are achieved. Together, the cases illustrate that abundance is not inherently beneficial or harmful, but rather that its creative value is determined by the degree of generative fit present in the collaboration. Because GenAI capabilities are evolving rapidly, readers should treat these cases as snapshots of fit or misfit identified in the 2023–2024 deployment conditions studied.

### *Scenario 1: Consulting Tasks at the Jagged Technological Frontier*

Dell'Acqua et al. (2023) conducted a pre-registered experiment with 758 consultants at Boston Consulting Group, randomly assigning them to one of three conditions: no AI access, GPT-4 access, or GPT-4 access with a prompt engineering overview. Participants performed two types of tasks: 18 creative and analytical consulting subtasks (e.g., ideation, market segmentation, persuasive writing) that fell inside AI's capability frontier, and a business case analysis task, requiring integration of qualitative interview data with spreadsheet reasoning, that fell outside it. On the frontier tasks, AI-using consultants completed 12.2% more tasks, worked 25.1% faster, and produced over 40% higher-quality output than the control group. Benefits were distributed across the skill spectrum but were largest for below-average performers (43% improvement vs. 17% for top performers). On the outside-the-frontier task, however, AI-using consultants were 19% less likely to reach the correct answer. The authors also identified two emergent integration patterns among successful AI users: "Centaurs," who strategically divided work between themselves and the AI, and "Cyborgs," who continuously interwove their workflow with the AI throughout the task. These results reveal sharp contrasts in generative fit across different task types.

For the 18 creative and analytical subtasks (ideation, market segmentation, writing), AI-using consultants achieved over 40% higher quality and 12% more task completion than control group (no AI). This demonstrates strong evocative fit: GPT-4's abundance of ideas, draft variants, and framings complemented consultants' ability to explore the solution space more broadly and quickly. The prompt engineering overview further enhanced fit by helping consultants extract better outputs, a form of organizational-level adaptive fit through training design.

On the business case requiring integration of qualitative interview data with spreadsheet analysis, AI-using consultants were 24% (GPT + Overview) and 13% (GPT only) less likely to reach the correct answer. This represents a failure of adaptive fit: the same unconstrained AI access was provided for a task requiring critical human judgment that the AI could not reliably perform, at least in 2023. Consultants who blindly adopted AI output exemplify the monoculture pathway: abundance produced a plausible-sounding answer that suppressed the exploratory skepticism needed to triangulate across data sources.

Even inside the frontier, semantic similarity analysis revealed that AI-assisted ideas were higher quality but more homogeneous across participants. This signals a partial failure of evocative fit at the social and organizational architecture levels: while individual exploration was enhanced, the collective diversity of the group's output narrowed because no workflow design encouraged consultants to diverge from AI's high-probability suggestions. The Centaur and Cyborg integration patterns observed among successful users suggest that open-ended fit can be partially recovered when individuals actively negotiate the human-AI boundary rather than defaulting to full delegation.

### *Scenario 2: Ad Copywriting with LLM Collaboration*

Chen and Chan (2024) conducted an experiment in which expert and non-expert participants wrote advertisement copies under three conditions: no AI assistance, an LLM used as a ghostwriter (generating draft ad copy), or an LLM used as a sounding board (providing feedback on human-written copy). Ad quality was measured by actual click performance on major social media platforms. The results diverged sharply by collaboration modality and user expertise. Non-experts using the LLM as a sounding board produced significantly higher-quality ads and closed the performance gap with experts; textual analysis showed their output converged semantically toward expert-level content, suggesting the LLM's feedback helped them explore marketing techniques they would not have reached independently. By contrast, when experts used the LLM as a ghostwriter, their performance declined relative to the no-AI control. Textual analysis revealed an anchoring effect: the LLM's initial drafts channeled experts toward generic, high-probability solutions, compressing rather than expanding their creative range. Ghostwriter-produced ads also exhibited the lowest semantic divergence across all conditions, indicating homogenization of output. Neither modality supported iterative revision or unanticipated recombination; the ghostwriter mode particularly reduced the extent to which participants revised their drafts, limiting the potential for discovery within the single-session workflow. These results reveal clear instances of good and poor generative fit across all three dimensions.

When non-experts used the LLM as a sounding board, they produced significantly higher-quality ads than those non-experts who did not have access to the LLM and closed the performance gap with experts. This exemplifies strong evocative fit at the cognitive architecture level: LLM feedback stimulated non-experts to explore marketing techniques they would not have considered independently, expanding their hypothesis space without supplanting their own generative process. The result was that non-experts' outputs became semantically closer to expert-produced ads, suggesting that abundance was channeled into productive exploration rather than overwhelming the user.

In the same experiment, when experts used the LLM as a ghostwriter, their performance declined significantly compared to the no-AI control. This illustrates a failure of adaptive fit: the same modality, that GenAI was prompted to compose ad copies, was applied regardless of expertise, but experts needed a different form of support than content generation. At the cognitive architecture level, the ghostwriter mode also violated evocative fit. Instead of stimulating exploration, the LLM's initial outputs anchored experts to high-probability, generic solutions, compressing rather than expanding the search space. Textual analysis confirmed that ghostwriter-produced ads exhibited the lowest semantic divergence across all conditions, a pattern similar to the one observed in Dell'Acqua et al. (2023) and, again, consistent with the monoculture pathway described in our framework: abundance existed upstream, but downstream selection collapsed around the most legible LLM-generated templates.

Across both modalities, the experiment reveals limited open-ended fit. Neither condition was designed to support emergence or unanticipated re-combinations over time. The ghostwriter mode particularly suppressed open-ended fit by reducing the extent to which participants revised their drafts, foreclosing iterative discovery. This organizational architecture constraint, a fixed, single-session workflow with no mechanism for serendipitous redirection, meant that even the sounding board's benefits were bounded by the structure of the task rather than the full potential of abundance.

### *Scenario 3: Skill-Biased AI-Augmented Creativity in Telemarketing*

Jia et al. (2024) conducted a field experiment and qualitative interview study at a telemarketing company where AI handled the initial, codified phase of work of generating sales leads, while human agents focused on the subsequent phase of creative sales persuasion. On average, AI assistance increased employee creativity in answering customer questions, which in turn boosted sales. However, the effect was sharply skill-biased. Higher-skilled agents benefited most: freed from repetitive lead generation and exposed to a concentrated stream of more serious customers, they developed innovative scripts, improvised new approaches, and reported positive emotions and a sense of creative freedom. The sequential division of labor matched AI's strength to the codified task phase and redirected skilled agents' cognitive resources toward the phase demanding creativity. Lower-skilled agents, by contrast, received the same intensified flow of challenging customers but lacked the domain knowledge to convert these harder cases into creative output. Rather than feeling freed to innovate, they experienced overload, anxiety, and demoralization, and wished for standardized answers instead. The identical AI deployment thus produced divergent outcomes depending on expertise, with no differentiated scaffolding or complementary training to help lower-skilled agents absorb the increased complexity.

When AI handled repetitive lead generation, top agents conserved cognitive resources and intensified exposure to challenging customers. This resulted in a form of abundance: more complex problems, and more opportunities to innovate telemarketing scripts. This achieved adaptive fit: the workflow matched AI's strength (codified, scripted work) to the task phase where it excelled, while redirecting human effort to the phase demanding creativity. It also enabled evocative fit: freed time and concentrated interactions with serious customers stimulated top agents to develop new scripts, reframe existing answers, and improvise on the spot using AI's output as a springboard for exploration rather than a finished product. Interviews revealed positive emotions, higher morale, and a sense of creative freedom, all indicating that abundance was converted into generative capacity. Both fitness enabled conserved resources used for deeper processing at cognitive level and the sequential division of labor as a deliberate job design choice that legitimized creative problem-solving for skilled agents at organizational level.

In contrast, the identical deployment produced a failure of fit for lower-skilled agents. Adaptive fit was absent because the workflow imposed a uniform design regardless of expertise: Low-skilled agents received the same intensified stream of challenging customers but lacked the domain knowledge of the higher-skilled agents to convert this abundance into creative output. Evocative fit failed because abundance in the form of harder cases did not stimulate exploration; instead, it triggered overload, anxiety, and demoralization. Rather than feeling freed to innovate, these agents wished for standardized answers, which is the opposite of generative exploration. Open-ended fit was also compromised: low-skilled agents could not discover emergent solutions through recombination because their limited absorptive capacity prevented them from building on the richer customer interactions. From the architecture angle, all three levels are impacted: cognitive (insufficient expertise to process increased complexity), social (negative emotions and reduced morale undermining collaborative dynamics), and organizational (absence of complementary training investments or differentiated scaffolding for varying skill levels).

In retrospect, this seemingly perplexing case demonstrates that abundance without expertise-calibrated generative fit produces skill-biased outcomes: Amplifying creativity for those already equipped to exploit it while intensifying pressure on those who are not thus equipped. Meanwhile, with deliberation calibration, generative fitness will be restored, as demonstrated in our next case (Luan et al. 2025).

### *Scenario 4: Augmented Learning in Human-GenAI Co-Creation*

Luan et al. (2025) conducted a three-study, mixed-methods investigation of human-GenAI co-creation in ideation tasks. In Studies 1 and 2, participants who collaborated with GenAI over multiple rounds did not automatically improve their joint creativity over time. Qualitative analysis of human-GenAI dialogues revealed the mechanism: participants increasingly defaulted to asking GenAI to produce idea lists without subsequent refinement, passively consuming abundant output rather than engaging with it. The decline in what the authors term "Idea Co-Development" or feedback exchanges and iterative idea refinement, was identified as the primary driver of creative stagnation. Despite GenAI's generative capacity, repeated reliance on the same prompting patterns caused dyads to converge on high-probability defaults rather than discovering novel directions. In Study 3, providing participants with structured instructions and guidance on Idea Co-Development restored creative improvement across rounds. When dyads exchanged critical feedback, reframed proposals, and iteratively refined ideas, GenAI's abundant output functioned as exploration stimuli rather than final answers, triggering new human associations and deeper elaboration. Dyads effectively shifted GenAI's role from wholesale generation to supporting the development phase where human input was most needed. Thus, abundance alone produced stagnation, while a deliberate organizational intervention, that is structured co-creation instructions, unlocked improved joint creativity.

Initially, when left to their own devices (Studies 1 and 2), human participants increasingly relied on asking GenAI to produce lists of ideas without subsequent refinement. Despite GenAI's abundant output, joint creativity flatlined. This illustrates a failure of all three fits. Evocative fit was absent because participants consumed outputs passively rather than using them as exploration prompts. Adaptive fit failed because the workflow did not shift from generation to evaluation as the bottleneck moved downstream – precisely the overload-to-monoculture pathway described in the abundance principle. Open-ended fit was lacking because repeated reliance on the same prompting pattern foreclosed emergence; dyads converged on GenAI's high-probability defaults rather than discovering unanticipated directions.

Then, when human-GenAI dyads engaged in idea co-development – exchanging critical feedback, reframing proposals, and iteratively refining ideas – they achieved improved joint creativity over rounds (Study 3). This new pattern exemplifies evocative fit: GenAI's abundant outputs served as stimuli for exploration rather than final answers, triggering new human associations and deeper elaboration. It also reflects adaptive fit: dyads effectively matched GenAI's generative capacity to the phase of work that most needed support (development rather than generation), flexibly calibrating each party's role to task demands.

This study's key intervention of providing structured collaborative co-creation instructions, functions as an organizational-level design choice that restored generative fit, demonstrating that abundance requires deliberate calibration rather than mere access.

## Discussion

This paper makes a creativity-first argument about human-GenAI collaboration. GenAI's distinctive contribution is not simply "better ideas," but abundance: low marginal cost generation of ideas, drafts, perspectives, critiques, and process actions at a pace that shifts the bottleneck from generation to attention, evaluation, selection, and integration. Organizing Acar et al.'s (2024) antecedents through a generativity lens, we argue that outcomes depend on generative fit as a configuration problem across cognitive, social, and organizational architectures.

### *Complexity and context-sensitivity as a feature, not a bug*

A key implication is that fit is not binary. It is continuous, multi-dimensional, and context-dependent, and it is precisely this complexity that gives the framework its analytical power because it reflects the structure of real creative work and explains why the same GenAI capability yields divergent outcomes across tasks, teams, and organizations.

### *The abundance-to-monoculture paradox as a theoretical contribution*

The paper's second discussion point is the apparent abundance-to-monoculture paradox. At first glance, abundance should increase diversity, since more outputs are possible at lower cost. Yet we often observe the opposite: convergence toward similar, high-probability solutions and reduced realized diversity. The framework resolves this paradox by locating it in a bottleneck shift. When GenAI produces far more inputs than teams can evaluate and integrate, the limiting resources become attention, selection, and integration capacity. Under scarcity, teams adopt coping moves that compress the search space, making monoculture a predictable outcome. The contribution is an explanatory mechanism: "more" can lead to "less" when abundance changes the constraints of the collaborative system. It also provides a practical diagnosis. When outputs become banal and overly similar, the issue is often not ideation effort but the configuration of evaluation, integration, and governance under abundance.

### *Extending generativity theory to human-GenAI collaboration*

A third theoretical contribution is that the framework extends generativity theory to human-GenAI collaboration by treating GenAI not as a passive tool, but as an active participant. It reshapes generative architecture, expands who can meaningfully participate, and changes the logic of governance and boundary resources, shifting the unit of analysis from individual outcomes to socio-technical systems that generate novelty through recombination, participation, and feedback over time. The paper also positions fit as the mechanism that converts abundance into outcomes. Outcomes depend on how architecture, community practices, and governance align to support exploration, task-appropriate calibration, and sustained emergence of new directions. This is particularly relevant because creative outcomes depend on both variation and integration (Acar et al., 2024), and generativity offers a system-level account of that coupling.

### *Practical implications: toward a diagnostic tool for managers and designers*

The framework has practical value as a diagnostic tool. When teams report that GenAI is not improving creativity, or outputs are repetitive, the framework can help ask "what is misaligned in our setup?" and pinpoint which form of fit is breaking down. Instead of focusing only on model capability, it focuses on configuration: whether the arrangement supports exploration and reframing (evocative fit), matches support to task phase and expertise (adaptive fit), and preserves room for emergence rather than locking work into a single template (open-ended fit). The framework guides targeted fine-tuning of architecture, community practices, and governance. Concretely, different task types call for different configurations. For novice-heavy ideation, favor sounding-board modalities and surface contrasting alternatives rather than a single "best" output (evocative fit). For expert convergent tasks, decouple generation from evaluation and avoid ghostwriting defaults that compress expert judgment (adaptive fit). For long-horizon or organizational creative work, preserve learning loops and boundary-spanning roles rather than standardizing GenAI output into templates (open-ended fit). These are starting heuristics; the broader point is that configuration choices are manipulable, making generative fit an actionable design target.

### *Future research directions*

The framework is designed to support empirical work. A near-term agenda is to operationalize the three fit dimensions into measurable constructs and manipulable design variables. Fit can be measured through survey items that capture perceived stimulation of exploration, perceived phase-sensitive alignment of support, and perceived preservation of emergence and renewal. Fit can also be manipulated experimentally, for example by varying whether systems present contrasting alternatives versus a single "best" output, whether evaluation is coupled or decoupled from generation, and whether workflows include structured critique and synthesis. A second direction is longitudinal and organizational. GenAI adoption can reshape community practices, learning loops, and boundary spanning over time. Studies that track how teams and organizations evolve their norms, evaluation routines, and governance, can test what increases monoculture risk and what preserves generative capacity. This also creates a bridge to research on skill development, deskilling, and the sustainability of human creative agency under abundant machine assistance.

## Conclusion

This paper argues that GenAI's defining contribution to creative collaboration is abundance: low marginal cost generation of many potentially relevant contributions across ideation, drafting, feedback, and process actions. Abundance can expand creative potential, yet it can also produce an apparent paradox, homogenizing convergence and reduced realized diversity. We explain this abundance-to-monoculture dynamic as a bottleneck shift: when generation becomes easy, attention, evaluation, selection, and integration become scarce, and teams sometimes cope by compressing the search space.

To explain when abundance becomes a creativity amplifier rather than a monoculture engine, we develop a generative fit view of human–GenAI collaboration. Fit is multi-dimensional and context-sensitive, and outcomes depend on how architecture, community practices, and governance align to support exploration, phase-appropriate calibration, and the emergence of unanticipated directions. The paper extends generativity theory by repositioning GenAI as an active participant in generative systems, and it offers a practical diagnostic for identifying why creativity gains fail to materialize and which configuration changes are likely to improve them.